\documentclass[a4paper, amsfonts, amssymb, amsmath, reprint, showkeys, nofootinbib, twoside]{revtex4-1}
\usepackage[english]{babel}
\usepackage[utf8]{inputenc}
\usepackage{amsthm}
\usepackage{mathtools}
\usepackage{xcolor}
\usepackage{graphicx}
\usepackage{adjustbox}
\usepackage{placeins}
\usepackage[T1]{fontenc}
\usepackage{lipsum}
\usepackage{csquotes}
\usepackage[dvipdfmx]{hyperref}
\usepackage{bm}
\begin{document}
\title{Geometrical Formulation of Adiabatic Pumping as a Heat Engine}

\author{Yuki Hino}
%\affiliation{Yukawa Institute for Theoretical Physics, Kyoto University, Kitashirakawa-oiwake cho, Sakyo-ku, Kyoto 606-8502, Japan}
\altaffiliation[Present Address: ]{NTT DATA Mathematical System Inc., 1F Shinanomachi 
Rengakan, 35, Shinanomachi, Shinjuku-ku, Tokyo 160-0016, Japan}

\author{Hisao Hayakawa}
    \email[Correspondence email address: ]{hisao@yukawa.kyoto-u.ac.jp}
\affiliation{Yukawa Institute for Theoretical Physics, Kyoto University, Kitashirakawa-oiwake cho, Sakyo-ku, Kyoto 606-8502, Japan}
%\affilation{Present Address: NTT DATA Mathematical System Inc., 1F Shinanomachi Rengakan, 35, Shinanomachi, Shinjuku-ku, Tokyo 160-0016, Japan}
\date{\today}

\begin{abstract}
We investigate a heat engine under an adiabatic (Thouless) pumping process. 
In this process, the extracted work and lower bound on dissipated availability are characterized by a vector potential and a Riemannian metric  tensor, respectively. 
We derive a trade-off relation between the power and effective efficiency. 
We also explicitly calculate the trade-off relation as well as the power and effective efficiency for a spin-boson model coupled to two reservoirs.
\end{abstract}
\maketitle

\section{INTRODUCTION}\label{sec:intro}

%Adiabatic pumping is a process where
 An average current can be generated even in the absence of an average bias under slow and periodic modulation of multiple parameters of the system. 
This is known as an adibatic pumping process.
Thouless first proposed the theory of the adiabatic pumping for an isolated quantum system \cite{thouless1,thouless2}. 
He showed that electrons can be transported by applying a time-periodic potential to one-dimensional isolated quantum systems under a periodic boundary condition. 
He also clarified that the charge transport in this system is essentially induced by a Berry-phase-like quantity in the space of the modulation parameters \cite{thouless1,thouless2,berry,xiao}.
This phenomenon has been observed experimentally in various processes such as charge transport \cite{ex-ch1,ex-ch2,ex-ch2.5,ex-ch3,ex-ch4,ex-ch5,ex-thou1,ex-thou2} and a spin pumping process  \cite{ex-spin1}.
Later, Brouwer extended the Thouless pumping to that in an open quantum system \cite{brouwer}. 
It has been  recognized that the essence of Thouless pumping can be described by a classical master equation in which the Berry-Sinitsyn-Nemenman (BSN) phase is the generator of the pumping current \cite{sinitsyn1,sinitsyn2}. 
There are various papers on geometrical pumping processes in terms of the scattering theory \cite{s-th1,s-th2,s-th3,s-th-ch1,s-th-ch2,s-th-ch3,s-th-spin1}, classical master equations\cite{parrondo,usmani,astumian1,astumian2,rahav,chernyak1,chernyak2,ren,sagawa} and quantum master equations \cite{qme1,qme2,qme-spin1,qme-spin2,yuge1,yuge2}. 
The extended fluctuation theorem for adiabatic pumping processes has also been studied \cite{watanabe,Hino-Hayakawa,Takahashi20JSP}. 

The geometrical concept is also used in finite time thermodynamics \cite{finite}, 
in which the thermodynamic length plays a key role.
The thermodynamic length is originally introduced for macroscopic systems \cite{Weinhold,Ruppeiner,Salamon,Schlogl,geo-rev}, and 
it has been used in wide classes of thermodynamic systems such as a classical nanoscale system \cite{Crooks}, a closed quantum system \cite{Deffner} and an open quantum system \cite{Scandi}.

Recently, Brandner and Saito have formulated the geometrical thermodynamics for a microscopic heat engine in the adiabatic regime \cite{Brandner-Saito}. 
In their approach, the properties of the working system are discribed by a vector potential and a Riemannian metric tensor in the space of control parameters. 
If one chooses a driving protocol, an effective flux and a length are assigned to the protocol. 
Then, they provide the extracted work and lower bound of the dissipated availability. 
On the other hand, Giri and Goswami proposed a quantum heat engine which includes the effect of the BSN phase by controlling temperatures of reservoirs \cite{engine}.

Nevertheless, we cannot apply the previous formulations to a heat engine undergoing an adiabatic pumping process with equal average temperature in both reservoirs, 
because (i) Ref. \cite{Brandner-Saito} only considered systems coupled to a single reservoir near an equilibrium state 
and (ii) Ref. \cite{engine} only considered the situation where the temperature of one of the reservoirs is always higher than that of the other one. 
Note that the outcome in the latter case is dominated by the dynamical phase. 
Namely, it is difficult to observe the geometrical contribution in such a system. 
In this paper, therefore, we extend the geometrical formulations of Refs. \cite{Brandner-Saito} and \cite{engine} to a system which is coupled to two reservoirs with equal average temperature under adiabatic modulation of thermodynamic quantities of the reservoirs and the target system. 
In the adiabatic regime, we obtain a geometrical representation of the extracted work and a lower bound of dissipated availability. We also derive a trade-off relation between the power and effective efficiency.

The organization of this paper is as follows. 
In Sec. \ref{sec:general}, we explain the setup and geometrical formulation for describing the heat engine under an adiabatic pumping process. 
In Sec. \ref{sec:appli}, we apply our formulation to a two-level spin-boson system coupled to two reservoirs.
Finally, we discuss and summarize our results in Sec. \ref{sec:conclusion}. 
%In Appendices, we present detailed calculations to support the description in the main text.
In Appendix A, we present a mathematical description of the pseudo-inverse of transition matrix. 
In Appendix B,  
we explain the outline of the perturbation theory of the master equation with slowly modulated parameters.
In Appendix C, we summarize the detailed setup of the spin-boson model used in the main text.

\section{GENERAL FRAMEWORK}\label{sec:general}

\subsection{Setup}\label{sec:setup}

\begin{figure}[htbp]
    \centering
\includegraphics[clip, width=7cm]{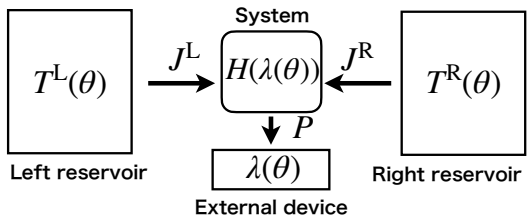}
    \caption{
%    {\color{red}{
    A schematic of the total system which consists of the target system and the left and right reservoirs. 
    We periodically control the temperatures of the reservoirs $T^{\mathrm{L}}$, $T^{\mathrm{R}}$ and the parameter $\lambda$ of the system Hamiltonian $\hat{H}(\lambda)$ by the external device. 
    $J^{\mathrm{L}}$ and $J^{\mathrm{R}}$ are the heat currents from the left and right reservoirs to the system, respectively.
    $P$ is the averaged power extracted from the system.
%    }}
    }
    \label{fig:engine}
\end{figure}

In this paper, we consider a system S coupled to two reservoirs L and R under periodic modulation of parameters with the period $\tau_{\mathrm{p}}$. 
A schematic of our system is depicted in Fig.~\ref{fig:engine}.
Each reservoir $\alpha=$L or R is characterized by the temperature $T^{\alpha}$.
We assume that the system S is characterized by $n$ discrete states $i=0, \dots, n-1$.
The system Hamiltonian $\hat{H}(\lambda)$ is characterized by its eigenvalue $E_i(\lambda)$ of $i-$th state with a control parameter $\lambda$.
 %$\hat{E}_{i}$ is the energy of $i$-th state and $\lambda$ is a control parameter. Here, 
%we have used the notation $\mathrm{diag}(\hat{E}(\lambda)$ for the diagonal matrix of $\hat{E}(\lambda)$, 
%in which the element $E_{ij}(\lambda)$ of $\hat{E}(\lambda)$ is expressed as $E_{ij}(\lambda)=E_i(\lambda) \delta_{ij}$. 
%Physically, $E_i(\lambda)$ is the energy of $i-$th state. 
%{\color{red}{
%For simplicity, we also use the notation $\mathrm{diag}(E_i)=\mathrm{diag}(\hat{E})$ for later discussion.}}
In this paper, we control the set of parameters $\bm{\Lambda} := (\lambda, T^{\mathrm{L}}, T^{\mathrm{R}})$.
% with period $\tau_{\mathrm{p}}$.

We assume that the dynamics of S is described by a master equation
\begin{align}\label{master}
    \frac{d}{d\theta} |p(\theta)\rangle 
    = \epsilon^{-1} \hat{K}(\bm{\Lambda}(\theta)) |p(\theta)\rangle.
\end{align}
Here we have introduced the dimensionless time (which is the phase in the modulation) $\theta := (t-t_{0})/\tau_{\mathrm{p}}$ and the dimensionless operation speed  $\epsilon := 1/(\tau_{\mathrm{p}} \Gamma)$, 
where  $t_{0}$ is the time after which the system reaches a periodic state and $\Gamma$ is the coupling strength or the characteristic transition rate between the system and the reservoirs.
%, which is proportional to the transition rate. 
%Note that $\Gamma$ is a model-dependent quantity.
The explicit expression of $\Gamma$ for the spin-boson model is presented in Appendix \ref{app:spin-boson}.
We have also introduced the vector 
$|p(\theta)\rangle := (p_{0}(\theta),\dots,p_{n-1}(\theta))^{T}$, where $p_{i}(\theta)$ is the probability of state $i$ at $\theta$.
The vector $|p(\theta)\rangle$ satisfies $0 \leq p_{i}(\theta) \leq 1$ and $\langle 1 | p(\theta) \rangle = \sum_{i} p_{i}(\theta) =1$, where $\langle 1 | := (1, \dots, 1)$.
The matrix $\hat{K}(\bm{\Lambda}(\theta))= \sum_{\alpha=\mathrm{L},\mathrm{R}} \hat{K}^{\alpha}(\bm{\Lambda}(\theta)) := \sum_{\alpha=\mathrm{L},\mathrm{R}}(k^{\alpha}_{ij}(\bm{\Lambda}(\theta)))$ is the transition matrix and its ($ij$)-component $k^{\alpha}_{ij}(\bm{\Lambda}(\theta))$ is the transition rate of $j\to i$ due to interaction with the reservoir $\alpha$ at $\theta$. 
$\hat{K}(\bm{\Lambda}(\theta))$ satisfies $\langle 1 | \hat{K}(\bm{\Lambda}(\theta)) = 0$.
We assume that the $\theta$-dependence of $\hat{K}(\bm{\Lambda}(\theta))$ only appears through the control parameters $\bm{\Lambda}(\theta)$.
We also assume that $\hat{K}^{\alpha}(\bm{\Lambda}(\theta))$ satisfies the detailed balance relation 
\begin{align}\label{DB}
    \ln \frac{k^{\alpha}_{ij}(\bm{\Lambda}(\theta))}{k^{\alpha}_{ji}(\bm{\Lambda}(\theta))} 
    = -\beta^{\alpha}(\theta) \left[ E_{i}(\lambda(\theta)) - E_{j}(\lambda(\theta)) \right],
\end{align}
where $\beta^{\alpha}({\theta}) := 1/T^{\alpha}({\theta})$ is the inverse temperature of the  reservoir $\alpha$ at $\theta$.
We assume that the master equation (\ref{master}) has a unique steady state $|p^{\mathrm{ss}}(\bm{\Lambda}(\theta)) \rangle$ which satisfies $\hat{K}(\bm{\Lambda}(\theta))|p^{\mathrm{ss}}(\bm{\Lambda}(\theta)) \rangle = 0$.
Since the system is coupled to two reservoirs having different temperatures,  $|p^{\mathrm{ss}}(\bm{\Lambda}(\theta)) \rangle$ is a nonequilibrium steady state.

\subsection{Thermodynamic Quantities}\label{sec:thermodynamic}

The performance of a heat engine is governed by the second law of thermodynamics, i. e. the non-negativity of the total entropy production rate.
When the system is coupled to a single heat  reservoir, the second law of thermodynamics is achieved by a quasi-static operation.
On the other hand, when the system is coupled to multiple heat  reservoirs, a proper {\it nonequilibrium} entropy production should be non-negative for arbitrary operations.
Such a non-negative quantity is known as the Hatano-Sasa (HS) entropy production rate $\sigma^{\rm HS}(\theta)$~\cite{Hatano-Sasa} defined as
%{\color {red}
\begin{align}\label{HS}
    \sigma^{\mathrm{HS}}(\theta) 
    := \dot{S}(\theta) + \sigma^{\mathrm{ex}}(\theta),
\end{align}
where
\begin{align}\label{Shannon}
    S(\theta) := \langle 1|\hat{s}(\theta) |p(\theta)\rangle
\end{align}
with its element $s_{ij}(\theta)$ of $\hat{s}(\theta)$ 
\begin{align}\label{Shannon_inside}
    s_{ij}(\theta) := -\ln p_{i}(\theta) \delta_{ij}
\end{align}
is the Shannon entropy of the system S with the notation $\dot{S}(\theta):=dS(\theta)/d\theta$.
Equation~\eqref{HS} contains the excess entropy production rate $\sigma^{\mathrm{ex}}(\theta)$ defined as
%{\color {red}
\begin{align}\label{excess_entropy}
    \sigma^{\mathrm{ex}}(\theta) := -\langle 1|\hat{\phi}(\bm{\Lambda}(\theta)) | \dot{p}(\theta)\rangle ,
\end{align}
%}
where the element $\phi_{ij}(\bm{\Lambda}(\theta))$ of $\hat{\phi}(\bm{\Lambda}(\theta))$ satisfies
\begin{align}\label{hat_phi}
 {\phi}_{ij}(\bm{\Lambda}(\theta)) := 
 -\ln p^{\mathrm{ss}}_{i}(\bm{\Lambda}(\theta))\delta_{ij} .
\end{align}
It is known that HS entropy production rate $\sigma^{\rm HS}$ is always non-negative and converges to zero in quasi-static limit $\epsilon\to 0$, thanks to the HS inequality \cite{Hatano-Sasa}. 

To discuss the performance of the heat engine, we introduce the dissipative availability \cite{Salamon, Brandner-Saito} defined as
\begin{align}\label{A-def}
 A := \int^{1}_{0} T(\theta) \sigma^{\mathrm{HS}}(\theta) d\theta,
\end{align}
where $T(\theta) := \beta(\theta)^{-1}$ with $\beta(\theta) := [\beta^{\mathrm{L}}(\theta) + \beta^{\mathrm{R}}(\theta)]/2$. 
According to HS inequality $\sigma^{\mathrm{HS}}(\theta) \geq 0$ the dissipative availability is always non-negative, i. e. $A \geq 0$.
Thus, $A$ plays a key role in nonequilibrium thermodynamics.
$A$ can be decomposed as
\begin{align}
 A = U + V - W,
\end{align}
where
\begin{align}\label{W}
 W := - \int^{1}_{0} \left\langle 1 \left| \frac{\partial \hat{H}(\lambda(\theta))}{\partial \lambda(\theta)} \right| p(\theta) \right\rangle \dot{\lambda}(\theta) d\theta
\end{align}
is the work extracted from the system per cycle.
Here, $U$ defined as
\begin{align}\label{U}
    U := - \int^{1}_{0} S(\theta) \dot{T}(\theta)d\theta 
\end{align}
is the thermal energy which can be used even in nonequilibrium processes~\cite{Brandner-Saito},
while $V$ defined as
\begin{align}\label{V}
    V := \int^{1}_{0} \left\langle 1 \left| \frac{\partial \hat{X}(\bm{\Lambda}(\theta))}{\partial \Lambda^{\mu}(\theta)} \right| p(\theta) \right\rangle \dot{\Lambda}^{\mu}(\theta) d\theta 
\end{align}
is a nonequilibrium potential which exists only if the system is coupled to multiple reservoirs.
Here we have introduced a matrix $\hat{X}(\bm{\Lambda}(\theta)) := T(\theta) [\hat{\phi}(\bm{\Lambda}(\theta)) - \hat{H}(\lambda(\theta))]$ and $\Delta(\theta) := [\beta^{\mathrm{L}}(\theta) - \beta^{\mathrm{R}}(\theta)]/2$.
Since $\bm{\Lambda}=(\lambda, T^{\mathrm{L}}, T^{\mathrm{R}})$ can be converted into $\bm{\Lambda}=(\lambda, T, \Delta)$, 
we use the notation $\Lambda^\mu$ to express one of $(\lambda, T, \Delta)$, i.e. $\Lambda^{0}=\lambda$, $\Lambda^1=T$ and $\Lambda^{2}=\Delta$ for later discussion.
%{\color{red} {
We also use Einstein's summation convention for $\mu=0, 1, 2$ in this paper.
%}}
Since $A \geq 0$, the extracted work $W$ is bounded as
\begin{align}
    W = U + V - A \leq U + V .
\end{align}
Thus, we can interpret $U + V$ as the available energy which can be converted into the work.
% and $A$ is the energy dissipation.
If a system is coupled to only one reservoir, the dissipative availability $A$ is reduced to $A=U-W$~\cite{Brandner-Saito}, which is equivalent to $W\leq U$.
Therefore, our dissipative availability $A$ is a nonequilibrium extension of the dissipative availability introduced in Refs.~\cite{Salamon, Brandner-Saito}.
  
Let us introduce the ratio of the work $W$ to the available energy $U+V$ as
\begin{align}\label{xi}
    \xi := \frac{W}{U+V} = \frac{W}{W+A}.
\end{align}
Because of $A\geq 0$, $\xi$ satisfies $0 \leq \xi \leq 1$. 
Thus, we call $\xi$ the effective efficiency because it is an indicator of the performance of the heat engine.
In the quasi-static limit ($\epsilon \to 0$, i. e. $A\to 0$), $\xi$ converges to $1$.
The scaled power defined as
\begin{align}\label{P}
    P := \epsilon W
\end{align}
converges to zero in this limit.
Note that $P$ does not have the dimension of power because we measure time scale by the dimensionless parameter $\epsilon$ under the fixing $\Gamma$.
To obtain larger power, we need the higher speed $\epsilon$ of operation, in which the effective efficiency $\xi$ becomes smaller.
In the next subsection, we discuss such a trade-off relation in the linear response regime.

We note that the conventional thermal efficiency is written as
\begin{align}\label{eta_def}
    \eta := \frac{W}{Q_{\mathrm{in}}},
\end{align}
where
\begin{align}
    Q_{\mathrm{in}} := \sum_{\alpha = \mathrm{L}, \mathrm{R}}\int^{1}_{0} d\theta \max [J^{\alpha}(\theta), 0]
\end{align}
is the heat absorbed by the system in one cycle and
\begin{align}
    J^{\alpha}(\theta) := \epsilon^{-1} \langle 1 |\hat{H}(\lambda(\theta)) \hat{K}^{\alpha}(\bm{\Lambda}(\theta)) |p(\theta)\rangle
\end{align}
is the heat current from the reservoir $\alpha$ to the system S at $\theta$.
If the temperature difference between two reservoirs is finite, the leading term of $J^{\alpha}(\theta)$ is $O(\epsilon^{-1})$. 
Thus, the thermal efficiency $\eta$ is $O(\epsilon)$, which converges to zero in the quasi-static limit. 

Let us briefly summarize difference between the effective efficiency $\xi$ and conventional efficiency $\eta$.
The former is the efficiency to express the conversion rate from the available source $U+V$ to the work $W$ once a nonequilibrium steady state is achieved, while the latter is the conversion rate from the absorbing heat to the work.
Both efficiencies prefer zero entropy productions to get high performance, but there are several intrinsic differences.
It should be noted that most of absorbing heat in a nonequilibrium engine is consumed as the housekeeping heat.
Therefore, $\xi$ is much higher than $\eta$ in general.
Moreover, the heat engine under Thouless pumping we consider in this paper is driven by reservoirs coupled to equal average temperature.
Therefore, $\eta$ becomes zero in the limit $\epsilon\to 0$.

%Let us rewrite $A$ as follows.
%{\color{red}{
For later discussion, let us rewrite the dissipative availability $A$.
Substituting Eqs.~\eqref{HS}-\eqref{hat_phi} into Eq.~\eqref{A-def} with the integral by part 
$A$ can be rewritten as
%{\color {red}{(You need to add some explanation, because $\int_0^1d\theta )\dot{T}\langle 1|(\hat{s}-\hat{\phi})|p\rangle$ seems not to be cancelled.}} 
\begin{align}\label{A_calc1}
	A =& \tilde{A}^{(0)}+\tilde{A}^{(1)}
\end{align}
where
\begin{equation}\label{cal_B}
\tilde{A}^{(0)}:= - \int^{1}_{0} \langle 1| [\hat{s}(\theta) - \hat{\phi}(\bm{\Lambda}(\theta))] |p(\theta) \rangle \dot{T}(\theta) d\theta
\end{equation}
and
\begin{align}\label{A-f}
    \tilde{A}^{(1)}: = - \int^{1}_{0} f^{\mu}(\theta) \dot{\Lambda}^{\mu}(\theta) d\theta
\end{align}
with
\begin{align}\label{f}
    f^{\mu}(\theta) := \langle 1| \hat{F}^{\mu}(\bm{\Lambda}(\theta))  |p(\theta)\rangle, 
\end{align}
and
\begin{align}    
    \hat{F}^{\mu}(\bm{\Lambda}(\theta)) 
    := - T(\theta)  \frac{\partial\hat{\phi}(\bm{\Lambda})}{\partial \Lambda^{\mu}}(\theta), 
\end{align}
which is the effective force conjugate to $\dot{\Lambda}^{\mu}(\theta)$.	
%	- \int^{1}_{0} d\theta 
%	\left\{ S(\theta) \dot{T}(\theta) 
%	- \left\langle 1 \left| \frac{d}{d\theta} \left( T(\theta) \hat{\phi}(\bm{\Lambda}(\theta)) \right) \right| p(\theta) \right\rangle \right\} 
%	\notag \\
%	&+ \left[ T(\theta) S(\theta) - T(\theta) \langle 1 | \hat{\phi}(\bm{\Lambda}(\theta)) | p(\theta) \rangle \right]^{1}_{0}
%	\nonumber
%	\\	
%	\notag \\
%	&+ \int^{1}_{0} \left\langle 1 \left| T(\theta) \frac{\partial \hat{\phi}(\bm{\Lambda})}{\partial\Lambda^{\mu}} (\theta) \right| p(\theta) \right\rangle \dot{\Lambda}^{\mu}(\theta) d\theta ,
%\label{A_calc2}
%\end{align}
To derive Eq.~\eqref{A_calc1} we have used the periodicities of $|p(\theta)\rangle$ and the control parameters $\bm{\Lambda}(\theta)$ to omit the boundary term. 
%{\color{red}

\subsection{Linear response regime}\label{sec:linear-response}

In this subsection, we consider thermodynamics of the engine introduced in the previous subsection in the linear response regime for small $\epsilon$. 
Thanks to Appendix \ref{app:slow-driving}, the solution of the master equation (\ref{master}) in the linear response regime can be expanded as
\begin{align}\label{p-ad}
    |p (\theta)\rangle \simeq |p^{\mathrm{ss}}(\bm{\Lambda}(\theta))\rangle + \epsilon |p^{(1)}(\bm{\Lambda}(\theta))\rangle +O(\epsilon^{2}),
\end{align}
where the second term on the right hand side of Eq. (\ref{p-ad}) can be written as
\begin{align}
    |p^{(1)}(\bm{\Lambda}(\theta))\rangle = \hat{K}^{+}(\bm{\Lambda}(\theta)) |\dot{p}^{\mathrm{ss}}(\bm{\Lambda}(\theta))\rangle.
\end{align}
Here $\hat{K}^{+}(\bm{\Lambda}(\theta))$ is the pseudo-inverse \cite{inverse} of the transition matrix $\hat{K}(\bm{\Lambda}(\theta))$ (see Appendix B for its details), which is written as
\begin{align}\label{pseudo-inverse}
    \hat{K}^{+}(\bm{\Lambda}) 
    = \int^{\infty}_{0} ds e^{s \hat{K}(\bm{\Lambda})} \Big( |p^{\mathrm{ss}}(\bm{\Lambda}) \rangle \langle 1 | - 1 \Big).
\end{align}
Then, $f^{\mu}(\theta)$ can be written as
\begin{align}\label{f-ad}
    f^{\mu}(\theta) \simeq f_{\mathrm{ss}}^{\mu}(\bm{\Lambda}(\theta)) + \epsilon R^{\mu\nu}(\bm{\Lambda}(\theta)) \dot{\Lambda}^{\nu}(\theta) + O(\epsilon^{2}),
\end{align}
where we have introduced %$f_{\mathrm{ss}}^{\mu}(\bm{\Lambda}(\theta))$ is defined as
%{\color{red}{
\begin{align}
    f_{\mathrm{ss}}^{\mu}(\bm{\Lambda}(\theta)) :=
     -\langle 1| \hat{F}^{\mu}(\bm{\Lambda}(\theta)) |p^{\mathrm{ss}}(\bm{\Lambda}(\theta))\rangle.
\end{align}
The response matrix $R^{\mu\nu}(\bm{\Lambda}(\theta))$ introduced in Eq.~\eqref{f-ad} is defined as
\begin{align}\label{R}
    R^{\mu\nu}(\bm{\Lambda}(\theta)) = -\beta(\theta) \int^{\infty}_{0} ds \langle F^{\mu}(s); F^{\nu}(0) \rangle_{\bm{\Lambda}(\theta)},
\end{align}
where
\begin{align}\label{cannonical}
    &\langle F^{\mu}(s); F^{\nu}(0) \rangle_{\bm{\Lambda}} 
    \notag \\
    &:=
    \left\langle 1 \left| \hat{F}^{\mu}(\bm{\Lambda}) e^{s \hat{K}^{+}(\bm{\Lambda})} \hat{F}^{\nu}(\bm{\Lambda}) \right| p^{\mathrm{ss}}(\bm{\Lambda}) \right\rangle 
    \notag \\
    &{~}- \left\langle 1 \left| \hat{F}^{\mu}(\bm{\Lambda}) e^{s \hat{K}^{+}(\bm{\Lambda})} \right| p^{\mathrm{ss}}(\bm{\Lambda}) \right\rangle
    \left\langle 1 \left| \hat{F}^{\nu}(\bm{\Lambda}) \right| p^{\mathrm{ss}}(\bm{\Lambda}) \right\rangle
\end{align}
is the cannonical correlation between $\hat{F}^{\mu}(\bm{\Lambda})$ and $\hat{F}^{\nu}(\bm{\Lambda})$ in a steady state characterized by $\bm{\Lambda}$.
Equation \eqref{cannonical} is nothing but the Green-Kubo formula in nonequilibrium systems coupled to two reservoirs.

Since we focus only on the linear region of $\epsilon$ in this paper, we can ignore $\tilde{A}^{(0)}$ in Eq. (\ref{A_calc1}), which is estimated as $O(\epsilon^{2})$.
\footnote{
This estimation can be shown as follows.
The integrand of the first term in $\tilde{A}^{(0)}$ in Eq.~\eqref{A_calc1} can be rewritten as
$\langle 1|(\hat{s}-\hat{\phi})|p\rangle
=\sum_i p_i \ln (p_i/p_i^{\rm ss})$, where we have used Eqs. \eqref{Shannon_inside} and \eqref{hat_phi}. 
Substituting Eq.~\eqref{p-ad} into this expression, we obtain
$\langle 1|(\hat{s}-\hat{\phi})|p\rangle
\simeq -\sum_i \{p_i^{\rm ss}+\epsilon p^{(1)} \}\ln \left\{1+\epsilon \frac{p^{(1)}}{p_i^{\rm ss}}\right\}+O(\epsilon^2)
=-\epsilon \sum_i p_i^{(1)}+O(\epsilon^2)$.
Because of $\sum_i p_i^{(1)}=0$ we obtain
$\langle 1|(\hat{s}-\hat{\phi})|p\rangle=O(\epsilon^2)$.
} 
%(HOW CAN WE ESTIMATE THIS?).
%}
Thus, Eq.~\eqref{A_calc1} is reduced to 
\begin{equation}\label{A=cal_A}
A\simeq \tilde{A}^{(1)} .
\end{equation}
Thus, substituting Eq.~\eqref{f-ad} into Eq.~\eqref{A-f} the dissipative availability $A$ can be rewritten as
\begin{align}\label{A-ad-0}
    A \simeq& 
%    A^{(0)}
    - \int^{1}_{0} f_{\mathrm{ss}}^{\mu}(\bm{\Lambda}(\theta)) \dot{\Lambda}^{\mu}(\theta) d\theta \notag \\ 
    &
    - \epsilon \int^{1}_{0} R^{\mu\nu}(\bm{\Lambda}(\theta)) \dot{\Lambda}^{\mu}(\theta) \dot{\Lambda}^{\nu}(\theta) d\theta 
    %+O(\epsilon^{2}),
\end{align}
where the first term 
%$A^{(0)}:=- \int^{1}_{0} f_{\mathrm{ss}}^{\mu}(\bm{\Lambda}(\theta)) \dot{\Lambda^{\mu}}(\theta) d\theta$ 
on the right hand side of Eq. (\ref{A-ad-0}) is zero because of
$\langle 1|d\phi(\theta)/d\theta|p^{\rm ss}\rangle =-\sum_ip_i^{\rm ss}
(d/d\theta)
\ln p_i^{\rm ss}=0
$
with the aid of $\frac{d}{d\theta}\sum_ip_i^{\rm ss}(\theta)=0$ and Eqs.~\eqref{A-f} 
and \eqref{A=cal_A}.
Thus, we obtain
\begin{align}\label{A-ad}
    A &= \epsilon \mathcal{A}_1+O(\epsilon^{2}) , \\
    \mathcal{A}_1:&= \int^{1}_{0} g^{\mu\nu}(\bm{\Lambda}(\theta)) \dot{\Lambda}^{\mu}(\theta) \dot{\Lambda}^{\nu}(\theta) d\theta ,
\end{align}
where
\begin{align}\label{metric}
    g^{\mu\nu}(\bm{\Lambda}(\theta)) := - \frac{1}{2} [R^{\mu\nu}(\bm{\Lambda}(\theta)) + R^{\nu\mu}(\bm{\Lambda}(\theta))]
\end{align}
is the thermodynamic metric tensor, which is symmetric and positive semi-definite.
Using the Cauchy-Schwartz inequality, the dissipative availability $A=\epsilon \mathcal{A}_1$ is bounded as
\begin{align}\label{CS}
   \mathcal{A}_1 \geq  \mathcal{L}^{2},
\end{align}
where
\begin{align}\label{L}
    \mathcal{L} := \oint_{\gamma} \sqrt{g^{\mu\nu}(\bm{\Lambda}) d\Lambda^{\mu} d\Lambda^{\nu}}
\end{align}
is the thermodynamic length corresponding to the length along the trajectory $\gamma$ in Riemannian manifold with the metric $g^{\mu\nu}(\bm{\Lambda})$.
The inequality \eqref{CS} is one of the main results in this paper, which has an identical form to that coupled to one reservoir~\cite{Brandner-Saito}.
The equality in Eq.~\eqref{CS} is held when $g^{\mu\nu}(\bm{\Lambda}(\theta)) \dot{\Lambda}^{\mu}(\theta) \dot{\Lambda}^{\nu}(\theta)$ is a non-negative $\theta$-independent constant. 
This equality cannot be achieved if BSN phase is meaningful, because $g^{\mu\nu}(\bm{\Lambda}(\theta)) \dot{\Lambda}^{\mu}(\theta) \dot{\Lambda}^{\nu}(\theta)$ should be a $\theta$-dependent variable once the trajectory in the parameter space makes a closed loop to generate BSN phase.
%Thus, the equality cannot be achieved if BSN phase is meaningful, because the trajectory in the parameter space cannot be a closed loop if $g^{\mu\nu}(\bm{\Lambda}(\theta)) \dot{\Lambda}^{\mu}(\theta) \dot{\Lambda}^{\nu}(\theta)$ is a non-negative $\theta$-independent constant.
%}}

The average power can be approximated as
\begin{align}\label{P-ad}
    P = \epsilon \mathcal{W} + O(\epsilon^2)
\end{align}
for small $\epsilon$, where
\begin{align}\label{W-ad}
    \mathcal{W} := \oint_{\gamma} \mathcal{A}^{\mu}(\bm{\Lambda}) d\Lambda^{\mu}
\end{align}
is the adiabatic work defined as the line integration of the thermodynamic vector potential 
\begin{align}\label{vector-potential}
    \mathcal{A}^{\mu}(\bm{\Lambda}) 
    := \lambda \frac{\partial}{\partial \Lambda^{\mu}} \left\langle 1 \left| \frac{\partial\hat{H}}{\partial \lambda} \right| p^{\mathrm{ss}}(\bm{\Lambda}) \right\rangle
\end{align}
along the trajectory $\gamma$ of parameter control \cite{Brandner-Saito}.
Note that $\mathcal{A}^{\mu}(\bm{\Lambda})$ corresponds to the BSN vector in adiabatic pumping processes~\cite{sinitsyn1,sinitsyn2}.

By using the equality (\ref{CS}), the effective efficiency $\xi$ (\ref{xi}) is written as
\begin{align}\label{xi_eq}
    \xi &= 1 - \epsilon \frac{\mathcal{A}_1}{\mathcal{W}} + O(\epsilon^{2}) . %\notag %    &\leq 1 - \epsilon \frac{\mathcal{L}^{2}}{\mathcal{W}} + O(\epsilon^{2}).
\end{align}
Using Eq.~\eqref{CS} the relation \eqref{xi_eq} can be rewritten as
\begin{equation}\label{xi_inequality}
1-\xi \ge \epsilon \frac{\mathcal{L}^2}{\mathcal{W}}=\epsilon^2 \frac{\mathcal{L}^2}{P} ,
\end{equation}
where we have used Eq.~\eqref{P} for the final expression.
This relation tells us that the decrement of the effective efficiency is bounded by the thermodynamic length $\mathcal{L}$ and $\mathcal{W}$ or the power $P$, which becomes smaller if $\mathcal{W}$ or $P$ is larger. 
If we regard $\epsilon$ as a control parameter, and thus $P$ as an independent variable from $\mathcal{W}$, one obtains the trade-off relation between the power and effective efficiency
\begin{equation}\label{trade-off2}
P\le \left(\frac{\mathcal{W}}{\mathcal{L}}\right)^2(1-\xi) 
\end{equation} 
in the limit $\xi\to 1$ which is equivalent to $\epsilon\to 0$.
This bound is identical to that for one reservoir~\cite{Brandner-Saito}.
The maximum slope in Eq.~\eqref{trade-off2} is determined by ${\mathcal{W}}/{\mathcal{L}}$, where $\mathcal{W}$ and $\mathcal{L}$ are the geometric quantities.
To optimize the performance of the engine we should
choose smaller under the condition $\mathcal{A}_1 \to \mathcal{L}^2$.

It is obvious that $\xi$ in Eq.~\eqref{xi_eq} becomes 1 and the power $P$ in Eq.~\eqref{trade-off2} becomes zero in the adiabatic limit ($\epsilon\to 0$).
This corresponds to the Carnot efficiency in the conventional thermodynamics.
Note that the conventional efficiency $\eta$ introduced in Eq.~\eqref{eta_def} tends to zero in the quasi-static limit ($\epsilon\to 0$) even if $\xi$ reaches the maximum value 1 because we need the house keeping heat, though we do not know how $\eta$ depends on $\epsilon$ in general.

\section{APPLICATION to Spin-boson system}\label{sec:appli}

In this section, we apply the general framework developed in previous section to the spin-boson model in which a single spin is coupled to two bosonic reservoirs (see Fig. \ref{fig:SB} for a schematic of our system).
Under the Born-Markov approximation, the system follows the master equation. 
Moreover, if we ignore the initial relaxation process, the off-diagonal elements of the density matrix of the system is negligible. 
Thus, the system can be regarded as a classical one.
Detailed formulation of the spin-boson system as a quantum system is given in Appendix \ref{app:spin-boson}.
In this section, we only consider the classical limit of this system.

\subsection{2-level classical spin-boson model}

%%%%%%%%%%%%%%%%%
\begin{figure}[htbp]
    \centering
	\includegraphics[clip, width=7cm]{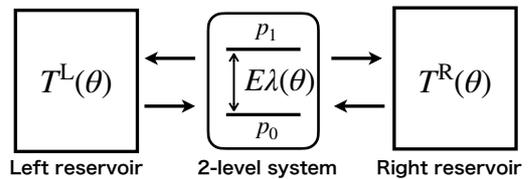}
    \caption{
%    {\color{red}{
    A schematic of a two-level spin-boson model.
    We periodically control the temperatures of the reservoirs $T^{\mathrm{L}}$, $T^{\mathrm{R}}$ and the parameter $\lambda$ of the system Hamiltonian $\hat{H}(\lambda)$. 
    $E\lambda(\theta)$ is the energy difference between two states.
%    }}
    }
    \label{fig:SB}
\end{figure}
%%%%%%%%%%%%%%%%%%%

The system contains one spin which has two states. The system Hamiltonian is given as
\begin{align}\label{H-SB}
    \hat{H}(\lambda(\theta)) 
    = \left(
    \begin{array}{cc}
        0 & 0 \\
        0 & E\lambda(\theta)
    \end{array}
    \right),
\end{align}
where $E\lambda(\theta)$ is the energy difference between two states with the non-negative control parameter $\lambda(\theta)$.
The system is coupled to two thermal reservoirs L and R characterized by temperatures $T^{\mathrm{L}}$ and $T^{\mathrm{R}}$, respectively.
We control the set of parameters $\bm{\Lambda}(\theta) = (\lambda(\theta), T^{\mathrm{L}}(\theta), T^{\mathrm{R}}(\theta))$ periodically with the control speed $\epsilon$.
The transition matrix $\hat{K}(\bm{\Lambda})$ of the master equation (\ref{master}) is given as
\begin{align}\label{K-SB}
    \hat{K}(\bm{\Lambda}(\theta))
    &= \left(
    \begin{array}{cc}
        - k_{10}(\bm{\Lambda}(\theta)) & k_{01}(\bm{\Lambda}(\theta)) \\
        k_{10}(\bm{\Lambda}(\theta)) & - k_{01}(\bm{\Lambda}(\theta))
    \end{array}
    \right) 
    \notag \\
    &= \sum_{\alpha} \left(
    \begin{array}{cc}
        - n^{\alpha}(\bm{\Lambda}(\theta)) & 1 + n^{\alpha}(\bm{\Lambda}(\theta)) \\
        n^{\alpha}(\bm{\Lambda}(\theta)) & - 1 - n^{\alpha}(\bm{\Lambda}(\theta))
    \end{array}
    \right),
\end{align}
where
\begin{align}
    n^{\alpha}(\bm{\Lambda}(\theta)) = \left( e^{\beta^{\alpha}(\theta)E \lambda(\theta)} - 1 \right)^{-1}
\end{align}
is the Bose distribution function in the reservoir $\alpha$ (= L, R).

The steady state of the master equation (\ref{master}) is given as 
\begin{align}\label{pss-SB}
    |p^{\mathrm{ss}}(\bm{\Lambda}(\theta)) \rangle
    &=
    \left(
    \begin{array}{c}
        p^{\mathrm{ss}}_{0}(\bm{\Lambda}(\theta)) \\
        p^{\mathrm{ss}}_{1}(\bm{\Lambda}(\theta)) 
    \end{array}
    \right) \nonumber\\
    &= \frac{1}{k_{01}(\bm{\Lambda}(\theta)) + k_{10}(\bm{\Lambda}(\theta))}
    \left(
    \begin{array}{c}
        k_{01}(\bm{\Lambda}(\theta))  \\
        k_{10}(\bm{\Lambda}(\theta))
    \end{array}
    \right).
\end{align}
In  $(\lambda, T, \Delta)$-representation, $|p^{\mathrm{ss}}(\bm{\Lambda}) \rangle$ can be written as
\begin{align}\label{2level-ss}
     p^{\mathrm{ss}}_{0}(\bm{\Lambda}(\theta)) 
    &= p^{\mathrm{eq}}_{0}(\lambda(\theta), \beta(\theta)) 
    \nonumber\\
    &{~}
    \left\{ 1 + k^{\mathrm{eq}}_{10}(\lambda(\theta), \beta(\theta))
     \Big[1 - \cosh{(\Delta E \lambda(\theta))} \Big] \right\}, \\
     p^{\mathrm{ss}}_{1}(\bm{\Lambda}) 
    &= p^{\mathrm{eq}}_{0}(\lambda(\theta), \beta(\theta)) 
    \nonumber\\
   &{~} \left\{ 1 + k^{\mathrm{eq}}_{01}(\lambda(\theta), \beta(\theta)) \Big[ \cosh{(\Delta E \lambda(\theta))} - 1 \Big] \right\},
\end{align}
where 
\begin{align}
    p^{\mathrm{eq}}_{0}(\lambda(\theta), \beta(\theta)) :&= \frac{1}{1 + e^{-\beta(\theta) E \lambda(\theta)}}, \\
    p^{\mathrm{eq}}_{1}(\lambda(\theta), \beta(\theta)) :&= \frac{e^{-\beta(\theta) E \lambda(\theta)}}{1 + e^{-\beta(\theta) E \lambda(\theta)}}
\end{align}
are the corresponding equilibrium states at $\beta$ for the ground state and excited state, respectively, (see Fig. \ref{fig:SB}) and 
\begin{align}
    &k^{\mathrm{eq}}_{10}(\lambda(\theta), \beta(\theta)) := 
    \left( e^{\beta(\theta) E \lambda(\theta)} - 1 \right)^{-1}, \\
    &k^{\mathrm{eq}}_{01}(\lambda(\theta), \beta(\theta)) := 
    e^{\beta(\theta) E \lambda(\theta)}\left( e^{\beta(\theta) E \lambda(\theta)} - 1 \right)^{-1}
\end{align}
are the transition rates at $\lambda(\theta)$ and $\beta(\theta)$.
Then, the diagonal elements of the matrix $\hat{X}(\bm{\Lambda}(\theta))$ in Eq. (\ref{V}) are explicitly given as
\begin{align}
    %\frac{
    X_{00}(\bm{\Lambda}) &= %-  \nonumber\\&{~}
   - T \ln \left\{ 1 + k^{\mathrm{eq}}_{10}(\lambda,\beta) \Big[1 - \cosh{(\Delta E \lambda)} \Big] \right\}, \\
    X_{11}(\bm{\Lambda}) &= - T % \nonumber\\  &{~}
    \ln \left\{ 1 + k^{\mathrm{eq}}_{01}(\lambda, \beta)) \Big[ \cosh{(\Delta E \lambda)} - 1 \Big] \right\} ,
\end{align}
where we have omitted writing $\theta$-dependence.

\subsection{Numerical calculation}

In this subsection, we calculate thermodynamic quantities discussed in Sec. \ref{sec:general} numerically. 
In this subsection, we control the set of parameters 
$\bm{\Lambda}(\theta) = (\lambda(\theta),T^{\mathrm{L}}(\theta), T^{\mathrm{R}}(\theta))$ as
\begin{align}
    \lambda(\theta) &= 1 + r_{\lambda} \cos[2\pi\theta], \\
    T^{\mathrm{L}}(\theta)/E &= c_{\mathrm{L}} + r_{\mathrm{L}} \sin[2\pi\theta], \\
    T^{\mathrm{R}}(\theta)/E &= c_{\mathrm{R}} + r_{\mathrm{R}} \sin[2\pi(\theta + \delta)],
\end{align}
where $\delta$ is the phase difference between the temperatures in left and right reservoirs.
We take $0 \leq \delta < 1/4$ without loss of generality. 
If we take $\delta \neq 0$, the temperature difference between two reservoirs remains finite. 
%Thus, $\delta$ characterizes the degree of nonequilibrium.
We also note that $\delta=0$ corresponds to the single-reservoir case discussed in Ref. \cite{Brandner-Saito}.

\begin{figure}[htbp]
    \centering
\includegraphics[clip, width=7cm]{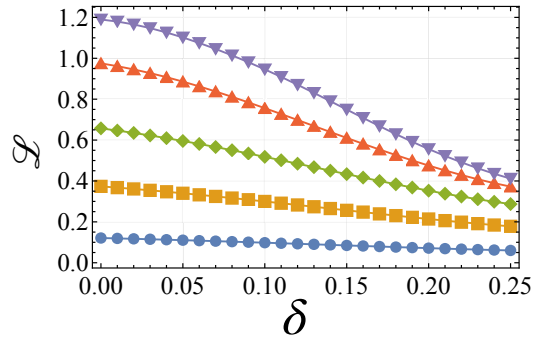}
    \caption{
        Plots of the $\delta$-dependence of the thermodynamic length $\mathcal{L}$ for  $\epsilon=0.01$, $c_{\mathrm{L}}=c_{\mathrm{R}}=1.0$, $r_{\lambda}=r_{\mathrm{L}}=r_{\mathrm{R}}=0.1, 0.3, 0.5, 0.7, 0.9$ corresponding to circle, square, diamond, up triangle and down triangle, respectively.
        }
    \label{fig:L}
\end{figure}

\begin{figure}[htbp]
    \centering
\includegraphics[clip, width=7cm]{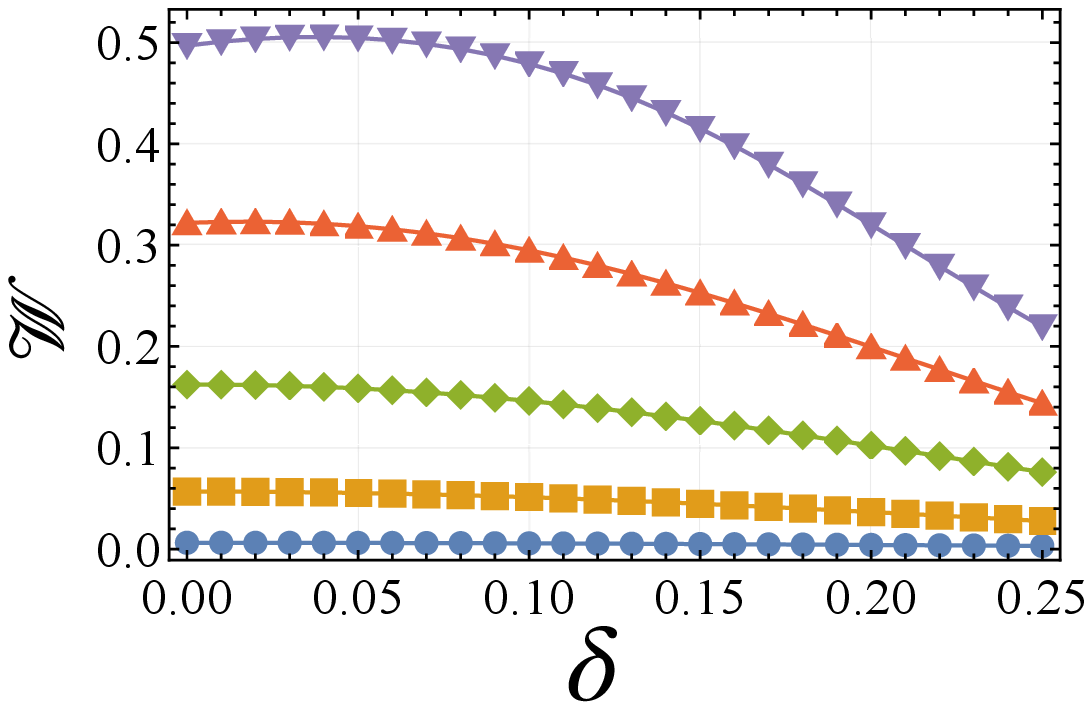}
    \caption{
        Plots of the $\delta$-dependence of the adiabatic work $\mathcal{W}$ for  $\epsilon=0.01$, $c_{\mathrm{L}}=c_{\mathrm{R}}=1.0$, $r_{\lambda}=r_{\mathrm{L}}=r_{\mathrm{R}}=0.1, 0.3, 0.5, 0.7, 0.9$ corresponding to circle, square, diamond, up triangle and down triangle, respectively.
        }
    \label{fig:W}
\end{figure}

\begin{figure}[htbp]
    \centering
\includegraphics[clip, width=7cm]{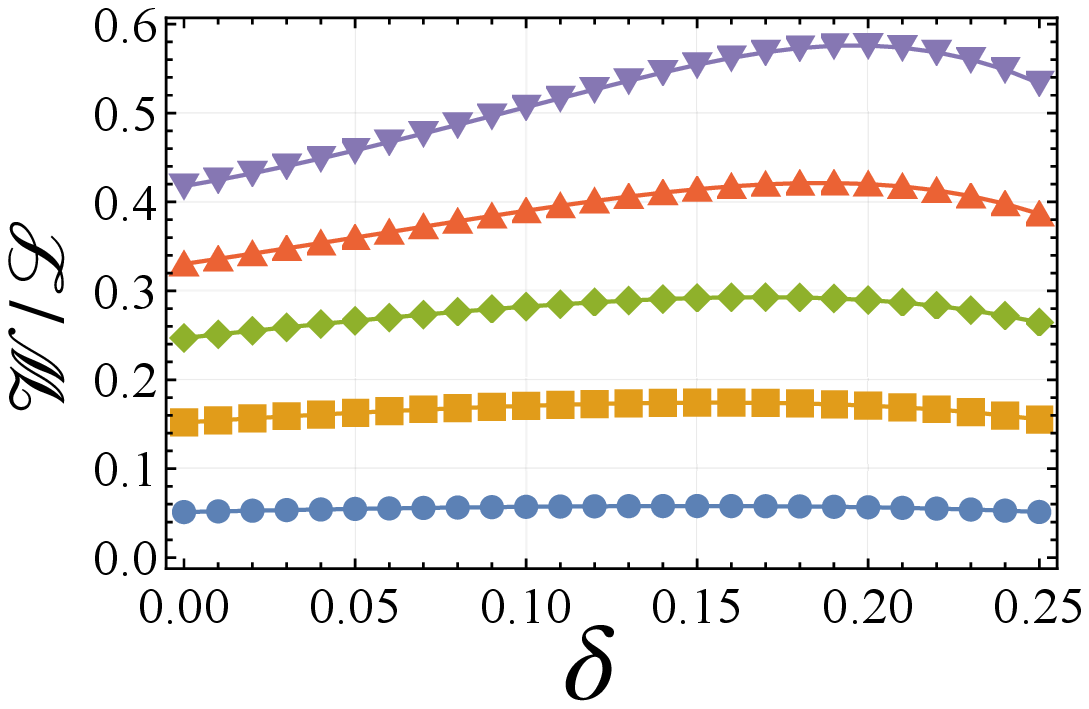}
    \caption{
        Plots of the $\delta$-dependence of the ratio $\mathcal{W}/\mathcal{L}$ for  $\epsilon=0.01$, $c_{\mathrm{L}}=c_{\mathrm{R}}=1.0$, $r_{\lambda}=r_{\mathrm{L}}=r_{\mathrm{R}}=0.1, 0.3, 0.5, 0.7, 0.9$ corersponding to circle, square, diamond, up triangle and down triangle, respectively.
        }
    \label{fig:WperL}
\end{figure}

\begin{figure}[htbp]
    \centering
\includegraphics[clip, width=7cm]{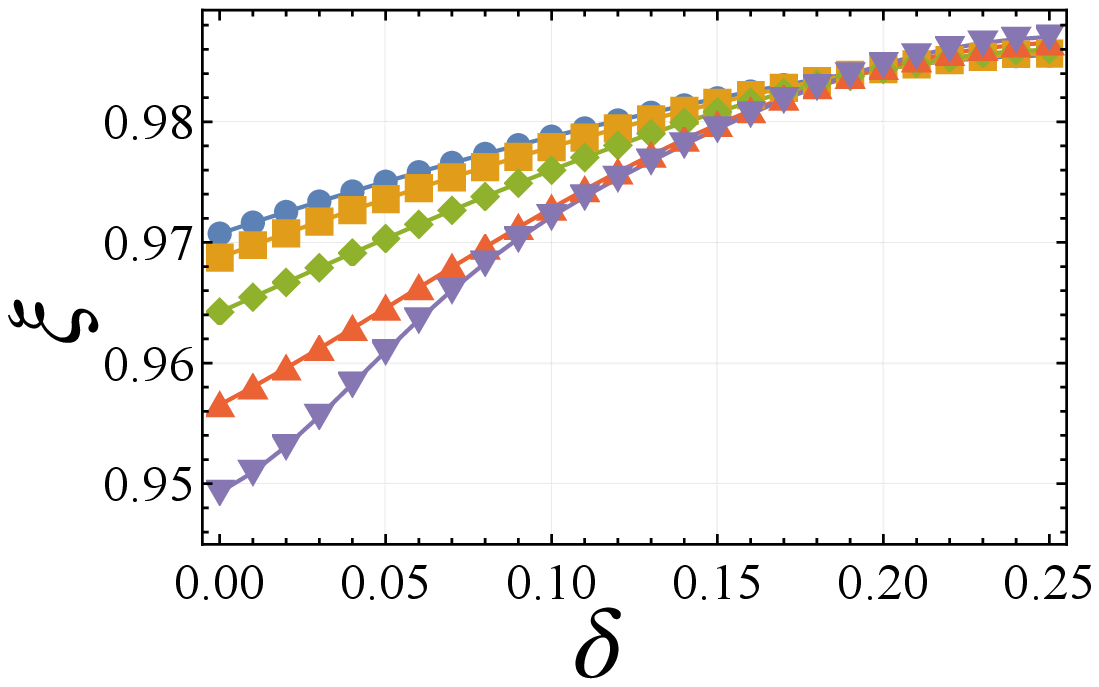}
    \caption{
        Plots of the $\delta$-dependence of the effective efficiency $\xi$ for $\epsilon=0.01$, $c_{\mathrm{L}}=c_{\mathrm{R}}=1.0$, $r_{\lambda}=r_{\mathrm{L}}=r_{\mathrm{R}}=0.1, 0.3, 0.5, 0.7, 0.9$ corresponding to circle, square, diamond, up triangle and down triangle, respectively.
        }
    \label{fig:xi}
\end{figure}

%As shown in Sec. \ref{sec:linear-response}, the ratio $\mathcal{W}/\mathcal{L}$ characterizes the performance of the heat engine. 
%Moreover,
We plot the $\delta$-dependences of the thermodynamic length $\mathcal{L}$, the adiabatic work $\mathcal{W}$, the ratio $\mathcal{W}/\mathcal{L}$  which plays an important role for the performance of the engine, and the effective efficiency $\xi$ in Figs. \ref{fig:L}, \ref{fig:W}, \ref{fig:WperL} and \ref{fig:xi}, respectively. 
The thermodynamic length $\mathcal{L}$ and the adiabatic work $\mathcal{W}$ monotonically decrease with $\delta$ (see Figs. \ref{fig:L} and \ref{fig:W}).
Thus, the geometric quantities such as $\mathcal{W}$ and $\mathcal{L}$ are a little suppressed by the heat current between two reservoirs.
The ratio $\mathcal{W}/\mathcal{L}$ takes a peak at a relatively large $\delta$ (see Fig. \ref{fig:WperL}).
The effective efficiency $\xi$ increases with $\delta$.
This relation suggests that we can make a high-performance engine if we ignore the house keeping heat to maintain a nonequilibrium steady state.

\begin{figure}[htbp]
    \centering
 \includegraphics[clip, width=7cm]{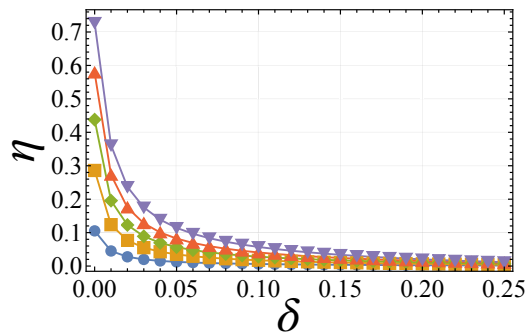}
    \caption{
        Plots of the $\delta$-dependence of the thermal efficiency $\eta$ for $\epsilon=0.01$, $c_{\mathrm{L}}=c_{\mathrm{R}}=1.0$, $r_{\lambda}=r_{\mathrm{L}}=r_{\mathrm{R}}=0.1, 0.3, 0.5, 0.7, 0.9$ corresponding to circle, square, diamond, up triangle and down triangle, respectively.
        }
    \label{fig:eta}
\end{figure}

We also plot the $\delta$-dependence of the conventional thermal efficiency $\eta$ in Fig. \ref{fig:eta}.
Contrary to $\xi$, the thermal efficiency $\eta$ decreases with $\delta$.
This is because $Q_{\mathrm{in}}$ contains the contribution of the steady heat current to maintain the nonequilibrium steady state, which increases with $\delta$.
This result is reaonable because the efficiency is expected to be high in a quasi-static operation near equilibrium.
If the system is far from equilibrium with increasing $\delta$,
we need the extra effort known as the house keeping heat  to maintain a nonequilibrium steady state. 

%%%%%%%%%%%%%%%%%%

\section{CONCLUSION}\label{sec:conclusion}

In this paper, we successfully extended the geometrical thermodynamics formulated in Refs. \cite{Brandner-Saito} and \cite{engine} to a system coupled to two slowly modulated reservoirs, i.e. the adiabatic (Thouless) pumping system without average bias. 
In the adiabatic regime, the extracted work can be written as the line integral Eq.~(\ref{W-ad}) of the thermodynamic vector potential Eq. (\ref{vector-potential})  along the path of the manipulation in the parameter space. 
On the other hand, the lower bound of the dissipated availability can be written as the thermodynamic length (\ref{L}) along the path. 
By using these results, we obtained the geometrical trade-off relation (\ref{trade-off2}) between the power and effective efficiency in the adiabatic limit. 
We applied these results to a two-level spin-boson system to obtain the explicit values of the power and effective efficiency. 
In contrast to Ref. \cite{engine}, we have analyzed a pumping system with two reservoirs of the same average temperature. 
Thanks to this setup, the geometrical contribution plays the dominant role in the thermodynamics of the heat engine. 

Our future tasks are as follows: 
(i) To calculate the thermodynamic metric or vector potential, we need to know the explicit form of the steady state of the master equation. 
In other words, our method cannot be used to systems for which a steady solution cannot be explicitly obtained.
Because nonequilibrium steady solutions cannot be obtained in most nonequilibrium systems, we need to extend our formulation even if the steady solution cannot be obtained.
(ii) The relationship between the effective efficiency $\xi$ and the conventional thermal efficiency $\eta := W/Q_{\mathrm{in}}$ is unclear in general systems. 
Clarifying this relationship is our future work.
(iii) In order to optimize the heat engine, one should find a trajectory that maximizes the ratio $\mathcal{W}/\mathcal{L}$ which could not be determined even in the spin-boson model in Sec. \ref{sec:appli}. 
The configuration of such an optimal trajectory is our future work.
(iv) Because the present method is restricted to the adiabatic case $\epsilon\to 0$, at least, for the argument after Sec. \ref{sec:linear-response}, we will need to extend the analysis to the non-adiabatic regime for finite $\epsilon$. 
Reference \cite{FHHT} obtained the non-adiabatic solution of a classical master equation and geometrical representation of the non-adiabatic current in two level system. 
We expect to apply these methods to investigate the non-adiabatic effect in heat engines. 
(v) Because we only focus on a classical system, we will have to try to extend our analysis to quantum systems in which quantum coherence plays an important role. 
Reference \cite{Brandner-Saito} showed that quantum coherence reduces the performance of slowly driven heat engines. 
On the other hand, it was shown that coherence can enhance the performance of heat engines in Ref. \cite{coherence}. 
Therefore, we will have to analyze full quantum systems to clarify whether the coherence can improve the efficiency in the heat engine undergoing an adiabatic pumping process.

\section*{Acknowledgements}
The authors thank Hiroyasu Tajima and Ken Funo for fruitful discussions. 
The authors also thank Ville Paasonen for his critical reading of this manuscript. 
This work is partially supported by a Grant-in-Aid of MEXT for Scientific Research (Grant No. 16H04025). 
The work of H.H. is partially supported by ISHIZUE 2020 of Kyoto University Research Development Program.

\appendix

\section{Slow-driving perturbation}\label{app:slow-driving}

%{\color {red}{Where is the place to cite this Appendix in the main text?}}
In this appendix, we explain the outline of the perturbation theory of the master equation with slowly modulated parameters \cite{slow_dynamics}. 
First, we expand the solution of Eq. (\ref{master}) in terms of $\epsilon$ as
\begin{align}\label{p-expand}
    |p(\theta)\rangle = \sum_{n=0}^{\infty} \epsilon^{n} |p^{(n)}(\bm{\Lambda}(\theta))\rangle.  
\end{align}
Since the normalization condition $\langle 1|p(\theta)\rangle =1$ holds for any $\epsilon$, 
$|p^{(n)}(\bm{\Lambda}(\theta))\rangle$ satisfies
\begin{align}
    &\langle 1|p^{(0)}(\bm{\Lambda}(\theta))\rangle = 1, \\
    &\langle 1|p^{(n)}(\bm{\Lambda}(\theta))\rangle =0 \;(n\geq 1).
\end{align}
Substituting these into Eq. (\ref{master}), we obtain
\begin{align}\label{p0-eq}
    &\hat{K}(\bm{\Lambda}(\theta))|p^{(0)}(\bm{\Lambda}(\theta))\rangle = 0, \\
    \label{pn-eq}
    &\hat{K}(\bm{\Lambda}(\theta))|p^{(n)}(\bm{\Lambda}(\theta))\rangle = \frac{d}{d\theta}|p^{(n-1)}(\bm{\Lambda}(\theta))\rangle \;(n\geq 1).
\end{align}
Equation (\ref{p0-eq}) means that $|p_{0}(\bm{\Lambda}(\theta))\rangle$ is the instantaneous steady state of $\hat{K}(\bm{\Lambda}(\theta))$: 
\begin{align}\label{p0}
    |p^{(0)}(\bm{\Lambda}(\theta))\rangle= |p^{\mathrm{ss}}(\bm{\Lambda}(\theta))\rangle.
\end{align} 
By using the pseudo-inverse $\hat{K}^{+}(\bm{\Lambda}(\theta))$ of $\hat{K}(\bm{\Lambda}(\theta))$, Eq. (\ref{pn-eq}) can be written as
\begin{align}\label{pn}
    |p^{(n)}(\bm{\Lambda}(\theta))\rangle 
    &= \hat{K}^{+}(\bm{\Lambda}(\theta)) \frac{d}{d\theta} |p^{(n-1)}(\bm{\Lambda}(\theta))\rangle \notag \\
    &= \left(\hat{K}^{+}(\bm{\Lambda}(\theta)) \frac{d}{d\theta} \right)^{n} |p^{\mathrm{ss}}(\bm{\Lambda}(\theta))\rangle.
\end{align}
Ignoring terms of $O(\epsilon^{2})$ and higher in Eq. (\ref{p-expand}), we obtain Eq. (\ref{p-ad}) of the main text.

\section{Pseudo-inverse of the transition matrix}\label{app:pseudo-inverse}

In this appendix, we introduce the pseudo-inverse $\hat{K}^{+}(\bm{\Lambda})$ of $\hat{K}(\bm{\Lambda})$, which satisfies following conditions \cite{inverse, Mandal}
\begin{align}
    \label{c1}
    &\hat{K}^{+}(\bm{\Lambda})\hat{K}(\bm{\Lambda}) = 1 - |p^{\mathrm{ss}}(\bm{\Lambda})\rangle \langle 1|, \\
    \label{c2}
    &\hat{K}(\bm{\Lambda})\hat{K}^{+}(\bm{\Lambda}) = 1 - |p^{\mathrm{ss}}(\bm{\Lambda})\rangle \langle 1|, \\
    \label{c3}
    &\hat{K}^{+}(\bm{\Lambda})|p^{\mathrm{ss}}(\bm{\Lambda})\rangle = 0, \\
    \label{c4}
    &\langle 1|\hat{K}^{+}(\bm{\Lambda}) =0.
\end{align}
In particular, if  $\hat{K}(\bm{\Lambda})$ is diagonalizable, $\hat{K}(\bm{\Lambda}) = \sum_{m} \phi_{m}(\bm{\Lambda}) |r_{m}(\bm{\Lambda})\rangle \langle l_{m}(\bm{\Lambda})|$, $\hat{K}^{+}(\bm{\Lambda})$ can be written as
\begin{align}\label{K+}
    \hat{K}^{+}(\bm{\Lambda}) = \sum_{m\neq 0} \phi_{m}(\bm{\Lambda})^{-1} |r_{m}(\bm{\Lambda})\rangle \langle l_{m}(\bm{\Lambda})|,
\end{align}
where $\phi_{m}(\bm{\Lambda})$ is the eigenvalue and $|r_{m}(\bm{\Lambda})\rangle$, $\langle l_{m}(\bm{\Lambda})|$ are the corresponding right and left eigenvectors of $\hat{K}(\bm{\Lambda})$. 
Here we note that $\phi_{0}(\bm{\Lambda})=0$, then $|r_{0}(\bm{\Lambda})\rangle = |p^{\mathrm{ss}}(\bm{\Lambda})\rangle$ and $\langle l_{0}(\bm{\Lambda})| = \langle 1|$.
Here we assume that these eigenstates do not degenerate.

As we mentioned in Sec. \ref{sec:linear-response}, the pseudo-inverse $\hat{K}^{+}(\bm{\Lambda})$ can be also written as Eq. (\ref{pseudo-inverse}). 
We can easily check that the form of the pseudo-inverse in Eq. (\ref{pseudo-inverse}) also satisfies the above conditions (\ref{c1}) - (\ref{c4}).

\section{Spin-Boson model}\label{app:spin-boson}

In this appendix, we summerize the detailed setup of the spin-boson model used in Sec. \ref{sec:appli}.

In the spin-boson model, the total Hamiltonian is given by $H_{\mathrm{tot}} = H_{\mathrm{S}} + \sum_{\nu=\mathrm{L}, \mathrm{R}}[H_{\alpha} + H_{S\alpha}]$. 
Each term is given by
\begin{align}
&H_{\mathrm{S}} = \frac{\hbar \omega_{0}}{2} \sigma_{z},
\\
&H_{\alpha} = \sum_{k} \hbar \omega_{k,\nu} b_{k,\alpha}^{\dagger} b_{k,\alpha},
\\
&H_{\mathrm{S}\alpha} = \hbar \sigma_{x} \otimes \sum_{k} g_{k,\alpha}( b_{k,\alpha} + b_{k,\alpha}^{\dagger}),
\end{align}
where $\hbar\omega_{0}$ is the energy gap between the two levels in the target system. 
$\sigma_{z}$ and $\sigma_{x}$ are Pauli operators, where $\omega_{k,\nu}$ is the angular frequency at wave number $k$ for the $\alpha$-th reservoir and $b_{k,\alpha}$ ( $b_{k,\alpha}^{\dagger}$) is the boson annihilation (creation) operator for the $\alpha$-th reservoir, respectively.
Here $g_{k,\alpha}$ is the coupling constant, which is related to the spectral density function $D_{\alpha}(\omega) := 2\pi \sum_{k} g_{k,\alpha}^{2} \delta(\omega - \omega_{k,\alpha})$.  
For later analysis, we use the line-width $\Gamma^{\alpha} = D^{\alpha}(\omega)$ which is independent of $\omega$.

We assume that the bosonic reservoirs are always at equilibrium. 
Thus, the density matrix of the $\nu-$th reservoir is expressed as $\rho_{\alpha}^{\mathrm{eq}}(\beta^{\alpha})=e^{-\beta^{\alpha} H_{\alpha}}/Z_{\alpha}$ at the inverse temperature $\beta^{\alpha}$, where $Z_{\alpha}=\mathrm{Tr}_{\alpha}[e^{-\beta^{\alpha}H_{\alpha}}]$.

The total density matrix $\rho_{\mathrm{tot}}(t)$ follows von-Neumann equation.
Under Born-Markov approximation, the reduced dynamics of the target system $\rho(t) := \mathrm{Tr}_{\mathrm{L,R}}[\rho_{\mathrm{tot}}(t)]$ can be describe by the Lindblad master equation \cite{Breuer}.
Moreover, the diagonal part of $\rho(t)$ is described by the master equation:
\begin{align}
    \frac{d}{dt}\left(
        \begin{array}{c}
            p_{0}(t)  \\
            p_{1}(t) 
        \end{array}
    \right)
    = \sum_{\alpha=\mathrm{L, R}}\left(
        \begin{array}{cc}
            -\Gamma^{\alpha} n^{\alpha} & \Gamma^{\alpha} (1 + n^{\alpha}) \\
            \Gamma^{\alpha} n^{\alpha} & -\Gamma^{\alpha} (1 + n^{\alpha})
        \end{array}
    \right)
    \left(
        \begin{array}{c}
            p_{0}(t)  \\
            p_{1}(t) 
        \end{array}
    \right),
\end{align}
where $n^{\alpha}$ is the Bose distribution function in reservoir $\alpha$ given by
\begin{align}
    n^{\alpha} := \frac{1}{e^{\beta^\alpha\hbar\omega_{0}} -1}.
\end{align}
By using $\theta := t/\tau_{\mathrm{p}}$ and $\epsilon := 1/\tau_{\mathrm{p}} \Gamma$ with $\Gamma:=\sum_{\alpha}\Gamma^{\alpha}/2$, we obtain the normalized master equation with (\ref{K-SB}).

In this model, we control $\beta^{\mathrm{L}}$, $\beta^{\mathrm{R}}$ and $\omega_{0}$.
In Sec. \ref{sec:appli}, we have used the notation $E := \bar{\omega}_{0} := \int^{1}_{0} d\theta \omega_{0}(\theta)$ and $\lambda(\theta) := \omega(\theta)/\bar{\omega}_{0}$.

\end{document}